\documentclass[10pt,aps,pra,twocolumn,superscriptaddress,amsmath,showpacs,nofootinbib]{revtex4-1}
\usepackage{amsmath} 
\usepackage{graphicx} 
\usepackage{amssymb}
\usepackage{amsfonts}
\usepackage[utf8]{inputenc}
\usepackage[T1]{fontenc}

\newcommand\extra[1]{}
\newcommand\ket[1]{\ensuremath{|#1\rangle}}
\newcommand\bra[1]{\ensuremath{\langle#1|}}

\begin{document}

\title{Efficient {A}mplification of {P}hotonic {Q}ubits by {O}ptimal {Q}uantum {C}loning}

\author{Karol Bartkiewicz}
\email{bark@amu.edu.pl}

\affiliation{Faculty of Physics, Adam Mickiewicz University,
PL-61-614 Pozna\'n, Poland}
\affiliation{RCPTM, Joint Laboratory of Optics of Palack\'y University
and Institute of Physics of Academy of Sciences of the Czech Republic,
17. listopadu 12, 772 07 Olomouc, Czech Republic }

\author{Anton\'{i}n \v{C}ernoch}
\affiliation{RCPTM, Joint Laboratory of Optics of Palack\'y University
and Institute of Physics of Academy of Sciences of the Czech Republic,
17. listopadu 12, 772 07 Olomouc, Czech Republic }

\author{Karel Lemr}
\affiliation{RCPTM, Joint Laboratory of Optics of Palack\'y University
and Institute of Physics of Academy of Sciences of the Czech Republic,
17. listopadu 12, 772 07 Olomouc, Czech Republic }

\author{Jan Soubusta}
\affiliation{Institute of Physics of Academy of Science of the Czech Republic, Joint Laboratory 
of Optics of PU and IP AS CR, 17. listopadu 50A, 77207 Olomouc, Czech Republic}

\author{Magdalena Stobi\'{n}ska}
\affiliation{Institute of Theoretical Physics and Astrophysics, University of Gda\'{n}sk, ul. Wita Stwosza 57, 80-952 Gda\'{n}sk, Poland}
\affiliation{Institute of Physics, Polish Academy of Sciences, Al. Lotnik\'{o}w 32/46, 02-668 Warsaw, Poland}

\begin{abstract}
We demonstrate that a phase-independent quantum amplifier of a polarization
qubit  is a complementary amplifier of  the heralded qubit amplifier
[N. Gisin, S. Pironio and N. Sangouard, \prl{\textbf{105}},
070501 (2010)].  It employs
the multi-functional cloner in $1\to2$ copying regime, capable of
providing approximate copies of qubits given by various probability
distributions, and is optimized for distributions with axial symmetry.
Direct applications of the proposed solution are possible in quantum
technologies, doubling the range where quantum information is coherently
broadcast. It also outperforms natural nonlinear amplifiers that use
stimulated emission in bulk nonlinear materials. We consider the amplifier
to be an important tool for amplifying quantum information sent via
quantum channels with phase-independent damping. 
\end{abstract}


\maketitle
\parskip=0pt

\section{Introduction}

Photons are the best long-distance information carriers, although 
transmission channels are inevitably lossy. However, overcoming the 
problem  of transmission loss for long-distance quantum 
communication is a challenging problem which seems not to have an 
apparent solution. This seriously limits the development of quantum technologies.   

\begin{figure}
\includegraphics[width=8.5cm]{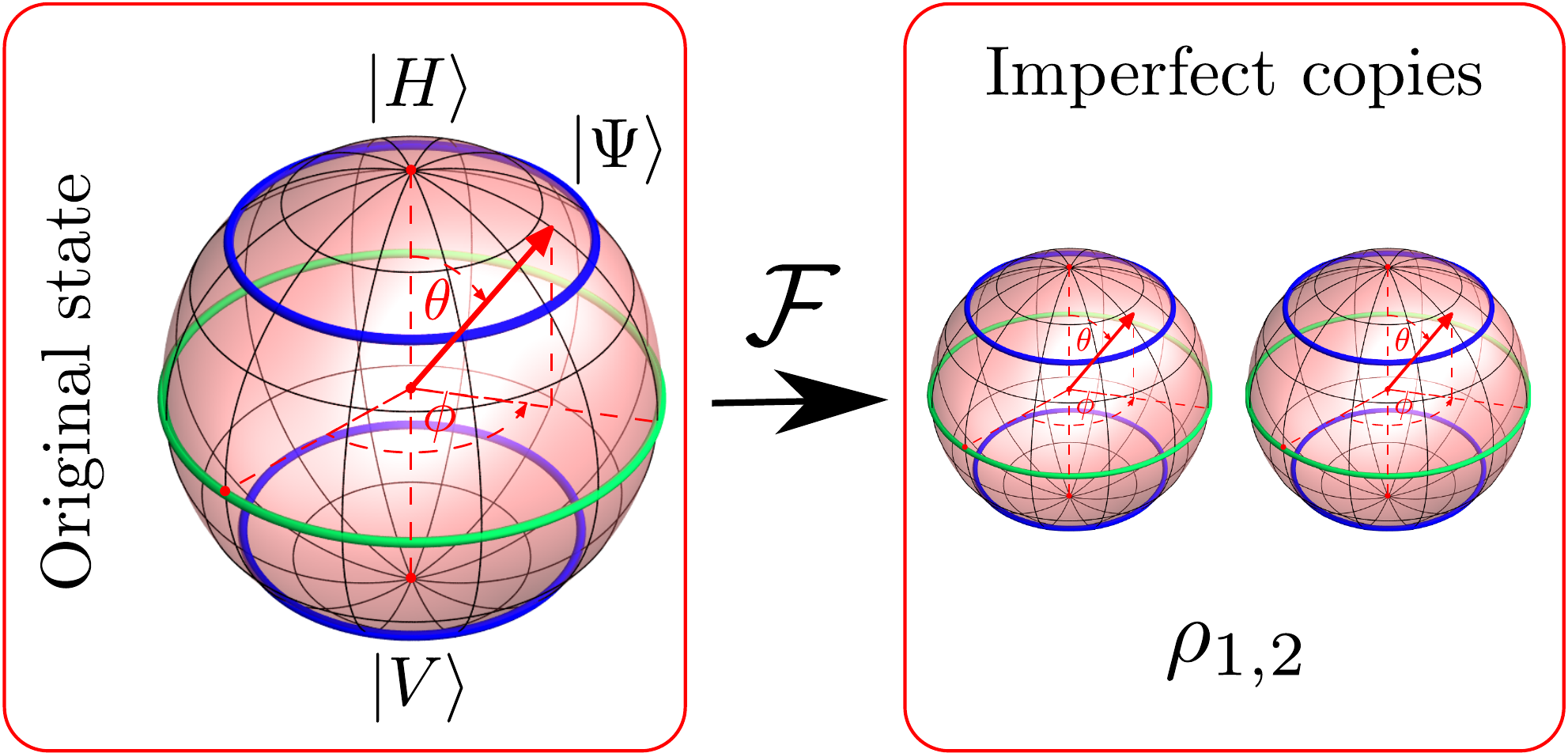} 
\caption{\label{fig:1}(Color online) Left: Poincaré sphere of 
polarization. A qubit is a linear combination of horizontal $|
H\rangle$ (the north pole) and vertical $|V\rangle$ (the south pole) 
polarizations: $|\Psi\rangle\!=\!\cos(\theta/2)|H\rangle\!+\! 
e^{i\phi}\sin(\theta/2)|V\rangle$. Radius of the sphere~is~$1$. The 
uniform probability distribution of polarization qubit is marked in 
red. Exemplary distributions with axial symmetry are in green and 
blue. Right: imperfect copies $\rho'$ of $\rho= 
|\Psi\rangle\langle\Psi|$.}
\end{figure}

There have been several proposal on how to increase the efficiency
of quantum information transfer including quantum repeaters 
(employing entanglement swapping \cite{Briegel98}  or quantum 
cloning \cite{Carlini03}) and heralded qubit amplifiers 
\cite{gisin10ampl}. A qubit is the smallest amount of quantum 
information represented by a vector in two dimensional Hilbert 
space.  Its natural optical implementation is the polarization of a 
single photon represented on the Poincaré sphere in Fig.~\ref{fig:1}
(left). Solving the problem of extending the range of quantum optical 
signals is essential for photonic quantum technologies: quantum 
communication, cryptography and metrology. Cloners and amplifiers
are useful not only for improving the effectiveness of quantum communication, they are also promising means of 
closing the detection loophole, paving the way to irrefutable Bell tests~\cite{Stobinska,Lvovsky} and device-
independent QKD \cite{gisin10ampl}.

Recently, a heralded noiseless amplifier of polarization qubits has
been proposed and \cite{gisin10ampl} and implemented \cite{Ralph10}. 
The heralded amplifier and its improvments (see, eg., Ref.~\cite{pitkanen11ampl})
or alternarive schemes using entanglement \cite{curty11ampl,Evan13,Bartkiewicz13sda}
have been demonstrated to be useful especially in the context of quantum key 
distribution. Interesingly the improved qubit amplifier described in
Ref.~\cite{curty11ampl} is closely related to entanglement swapping, 
which suggests that there is a strong connection between 
qubit amplifiers and quantum repeates (or relays), which are both based 
on similar underlying ideas. The heralded 
amplifier announces the time of arrival of optical 
signals. By doing so it increases the probability of receiving the 
signal after it was announced. This means that we are only waiting 
for a photon once it has been announced. Thus, this approach could
spare us some time otherwise wasted on waiting for the 
unannounced signal. The amplifier works with an arbitrary gain  
(increase in of the probability of registering a photon) without 
introducing noise, but the probability of successfully 
heralding event is decreases as the gain is increased. 
In the case of the originall proposal, perfect gain could have
been reached only asymptotically, however this shortcomming
has been removed by Pitkanen {\it et al.}~\cite{pitkanen11ampl}.
Recently, it has been shown that, if a heralded amplifier is allowed
to introduce some noise, its gain and herading rate be 
increased~\cite{Bartkiewicz13sda} which also depends on the set of qubits
to be amplified. 

Instead of increasing the probability of receiving the signal by
announcing its expected arrival one can use an alternative strategy 
based on sending multiple copies of the signal through the lossy 
channel. This means that we will receive a message faster, because 
there is no reason for repeating it once it was sufficiently amplified.
Similarly as before this approach can save us some time.
There is however, a fundamental problem with this approach
when we try to apply it in case of single photons in unknown 
quantum states.  In classical world the weak signal is analyzed and 
then a stronger version of it is produced and send 
further. This allows us to make the signal arbitrarily strong
which makes it more robust against transmission losses.
However, in quantum regime measuring the signal in a wrong basis 
will introduce errors. Moreover, according to the no-cloning 
theorem~\cite{Zurek82}, quantum  information cannot be perfectly 
copied or amplified. 

The best copying results are achieved by optimal quantum
cloning machines (QCMs)~\cite{Hillery,cloning}. They are crucial
for analyzing the security of quantum key distribution (QKD) against
coherent and incoherent attacks~\cite{attacks,Bartkiewicz13}, and
limit the capacity of quantum channels~\cite{Cerf2000}. They
provide the highest fidelity $\mathcal{F}$ between the original qubit
and its copies, see Fig.~\ref{fig:1}(right).
Universal QMCs operate equally well for all qubits, with 
$\mathcal{F}=5/6$ for two copies ($1\to2$)~\cite{Hillery}. State-
dependent cloners can exceed this fidelity for some subspaces of the 
sphere~\cite{cloning} using partial information about the input qubit 
distribution. Examples include the phase-covariant 
cloner~\cite{Bruss2000,cloning} optimal for
equatorial states; and its generalizations to phase-independent
cloners~\cite{gisin97NMcloner,Fiurasek2003,Hu2009,Bartkiewicz09,Bartkiewicz10,Lemr12},
optimal for qubits with axially-symmetric distributions, e.g., 
Fisher~\cite{Fisher}, Brosseau~\cite{Brosseau}, and Henyey-
Greenstein~\cite{Henyey}  distributions. These distributions are used 
to describe many problems of directional statistics ~\cite{dirstat} 
in physics, astronomy, biology, geology and psychology. 

Can we construct a useful qubit amplifier that will make use of the classical repetition scheme by using the best approximate quantum 
cloning operations?  To answer this question let us consider an erasure channel. If we replace each of the deleted qubits with a random state, we deal with a depolarizing channel. According to HSW (Holevo-Schumacher-Westmoreland) theorem~\cite{QSTbook}, depolarizing channel $\mathcal{E}: \rho\to \eta \rho + (1-\eta)\openone/2$, where $\rho$ is a pure state, yields product state capacity of
\begin{equation}
\label{eq:psc}
C(\mathcal{E}) = 1 - H\left(\frac{1+\eta}{2}\right),
\end{equation}
where $H(x) = -x\log_2x -(1-x)\log_2(1-x)$ is binary entropy. If we prepare two clones of the pure input state $\rho$ with fidelity $F$ and then send them both through the channel, the effective transformation (for small $\eta$) turns out to be  $\mathcal{E}': \rho\to \eta '\rho + (1-\eta')\openone/2$, where $\eta'=2(2F-1)\eta$ (for details see Sec.~II ). If $4F-2>1$, we can increase product state capacity by applying quantum cloning. In particular, this is true for universal cloning~\cite{Hillery}, where $F=5/6$, but not for classical cloning for which $F=3/4$. Therefore, we expect that information can be transmitted more efficiently if amplified by a quantum copying machine which provides fidelity $F>3/4$, i.e., $C(\mathcal{E}')>C(\mathcal{E})$. This is holds under assumption that the cloning operation is deterministic or its success probability is sufficiently large, i.e., $P > 1/(4F-2)$. Note, that there is no reason for the cloning process to work with $P<1$ other than the implementation-dependent technical issues. However, if we limit ourselves to the framework of linear optics, can such amplifier perform better, in terms of gain or qubit fidelity, than other types of qubit amplifiers?  

\begin{figure}
\includegraphics[width=8.5cm]{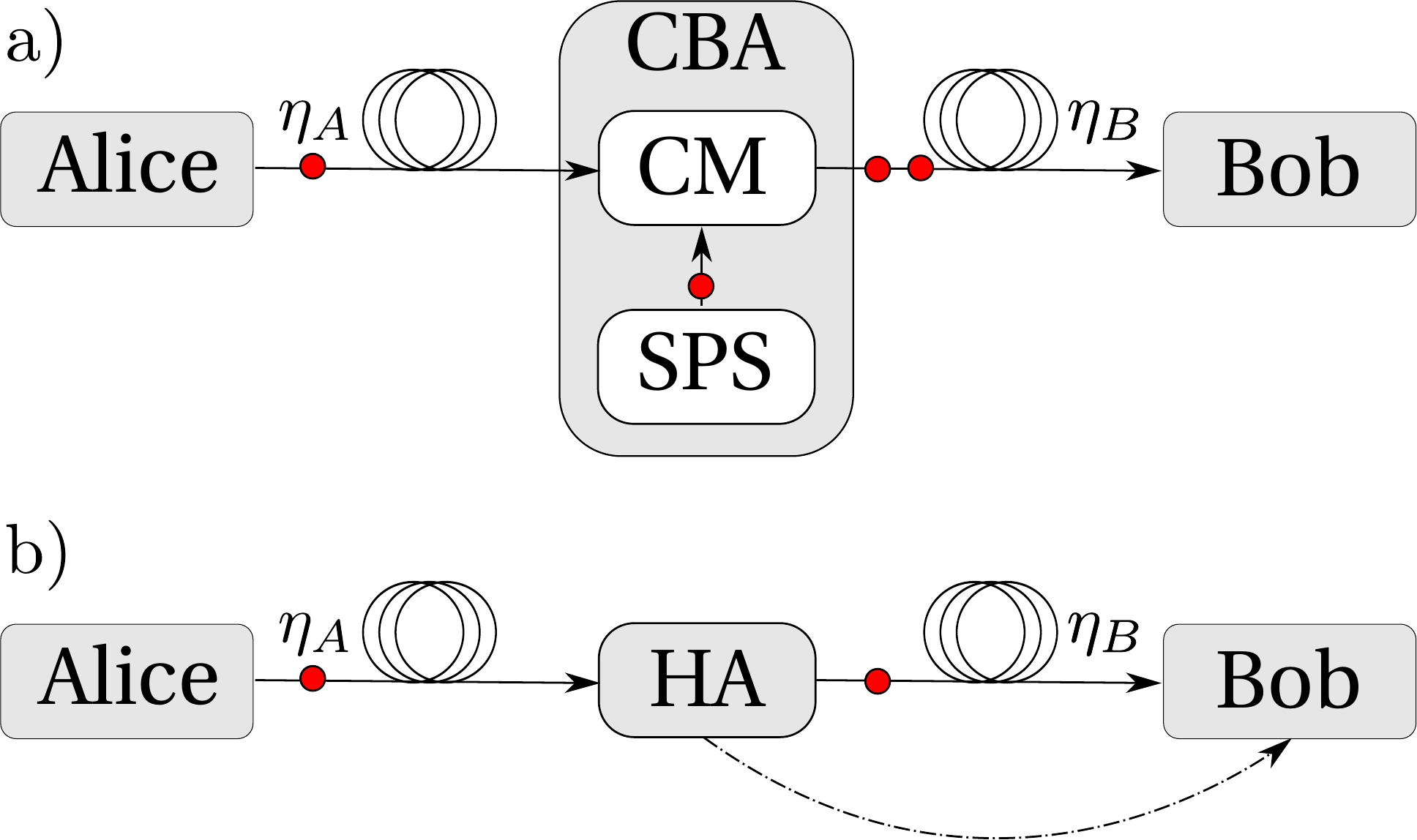} 
\caption{\label{fig:concept} (Color online) Both cloning-based amplifier (a) and heralded amplifier 
can be placed in the communication channel of total transmissivity $\eta = \eta_A\eta_B$ to fight the 
lossed. We assume that Alice sends a photon to Bob who wants to receive the quantum message with minimum time 
spent on waiting. Because of the channel losses only a fraction of photons will reach Bob. If one uses a qubit amplifier, probability of Bob receiving the signal is increased. In case of the CBA this probability is increased by sending an additional phtoton generated by a single photon source (SPS). If the HA is used, the probability is increased, by informming Bob when he should be ready to receive a photon so that he does ont waist time on waiting in vain.}
\end{figure}

We expect that cloning-based amplifier (CBA) will be directly applicable to quantum technologies by increasing the range over which quantum information is coherently broadcast. It will also outperform natural nonlinear 
amplifiers that use stimulated emission from bulk nonlinear 
materials. In this paper we will study cloning-based amplifiers and 
compare them with heralded amplifiers in terms of qubit fidelity and 
gain in probability of receiving the amplified signal for various 
communication scenarios (see Fig.~\ref{fig:concept}).
We will demonstrate that in some cases a cloning-based  phase-
independent quantum amplifier of a polarization qubit  outperforms 
the heralded qubit amplifier~\cite{gisin10ampl} (HA) in terms of the 
provided gain. However, as we show, the CBA performs worse than 
HA if used at the end of the transmission line. Thus, we conclude 
that the two types of amplifiers perform complementary tasks.
The CBA employs the $1\to2$ multi-functional 
cloner~\cite{Lemr12,Bartkiewicz10} optimized for distributions with 
axial symmetry. 

Some universal and phase-covariant QCMs are built using parametric
amplifiers and can  generate macroscopic quantum states of 
light~\cite{MacroRMP}. Other implementations use two-
photon interference on a polarization sensitive beam 
splitter~\cite{Lemr12}. Here we will use the latter approach to 
experimentally demonstrate a principle of operation of a CBA.

\section{Cloning based amplifier}

\begin{figure}
\includegraphics[width=8.5cm]{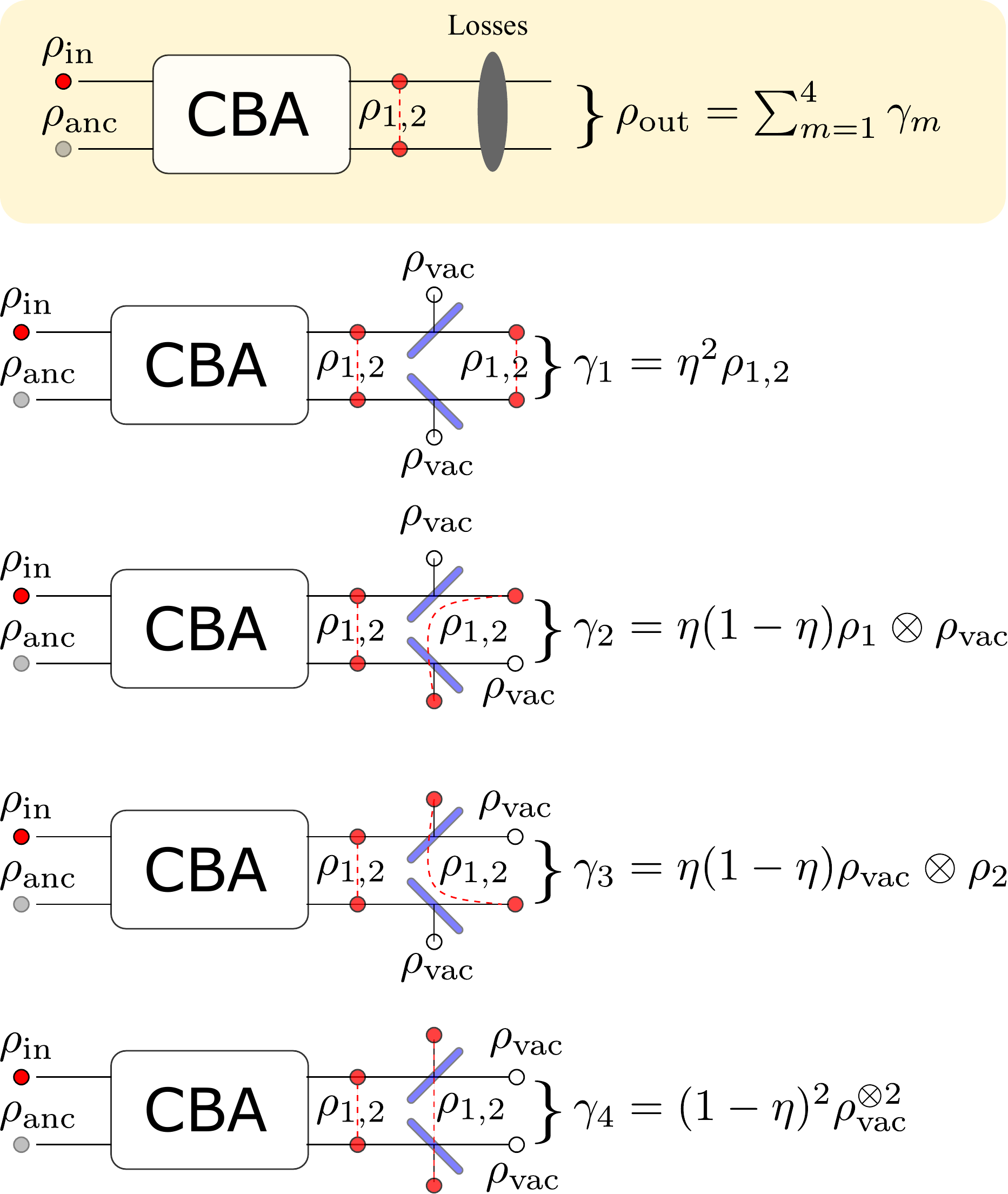} 
\caption{\label{fig:3}(Color online) Principle of operation of the CBA 
working at maximum efficiency $P=1$. The CBA has two input modes for signal 
qubit $\rho_{in}$ and ancillary qubit $\rho_{anc}$. The output of the CBA is 
described by state $\rho_{1,2}$. The reduced density matrices $\rho_1= 
\mathrm{Tr}_2 \rho_{1,2}=\rho'$ and $\rho_2=\mathrm{Tr}_1 \rho_{1,2}=\rho'$ 
describe each of the two copies of $\rho_{in}=\rho$ separately. Here we 
demonstrate how the output state of the amplifier is calculated by modeling 
losses as beam splitter transformations.  
}
\end{figure}

The CBA preamplifies a qubit  through duplication before sending its two 
copies via a lossy channel (see Fig.~\ref{fig:3}). One has to feed it with both 
the amplified qubit and an ancillary qubit. The total probability of finding at least one
photon at the output of the CBA working with efficiency $P=1$ is 
$P_{n>0}^{(\mathrm{CBA})}=1-(1-\eta)^2=(2-\eta)\eta$  (see Fig.~\ref{fig:3}) which is 
greater than the analogous probability for a bare channel, 
where $P^{(0)}_{n>0}=\eta$. Assume a photon reaches its destination with probability
$\eta$. If two photons are sent, the probability that at least one of them 
arrives is $2\eta(1-\eta)+\eta^{2}$, i.e., transmission efficiency increases.
If the CBA works with efficiency $P\leq1$ we have 
$P_{n>0}^{(\mathrm{CBA})}=P(2-\eta)\eta$. Let us define transmission gain for the CBA as
\begin{equation}
G^{(\mathrm{CBA})}_{T}(\eta,P) = P\frac{P_{n>0}^{(\mathrm{CBA})}}{P_{n>0}^{(0)}} = (2-\eta)P.
\label{eq:gain}
\end{equation} 
In contrast to HA, the CBA provides gain limited by the number of copies. The 
device provides a higher gain if $1\to N_{\mathrm{copy}}$ 
cloning~\cite{gisin97NMcloner,sciarrino07pcc}
is applied, although the clones have lower  fidelity. There are no laws of physics 
preventing to reach $P=1$, however, this value depends on the particular platform of 
implementing the amplifier. In the following sections we will focus on the case of 
$\eta\ll1$, where $G^{(\mathrm{CBA})}_{T}(P) = 2P \equiv G$.

\begin{figure}
\includegraphics[height=7.25cm]{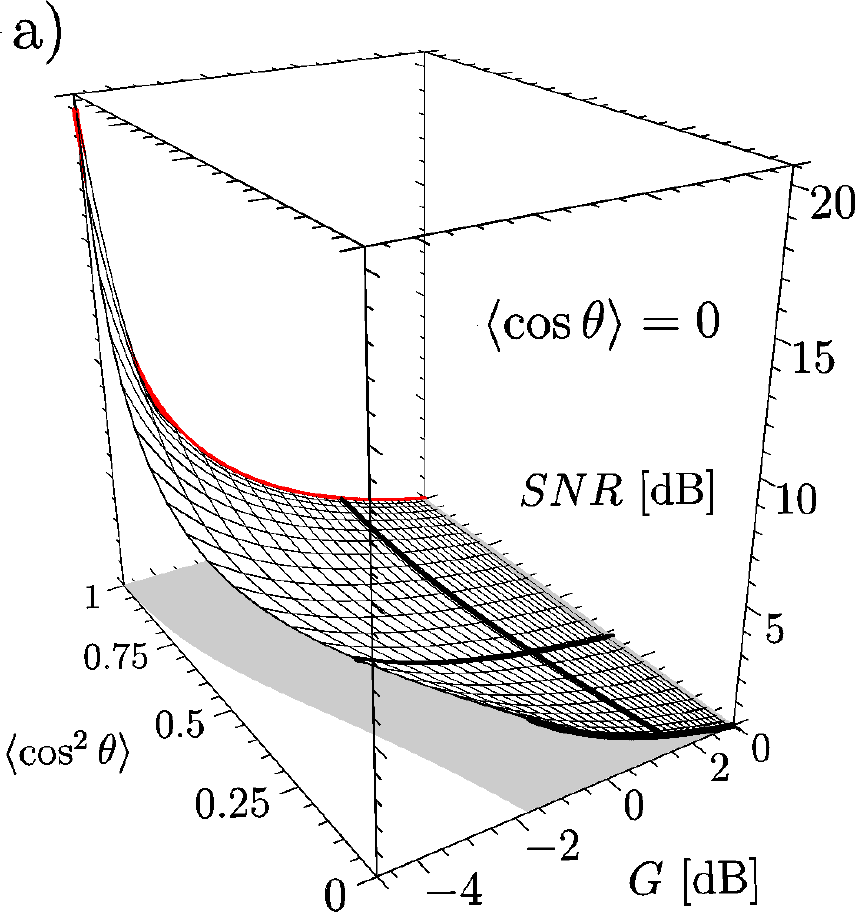} 
\vskip0.5cm
\includegraphics[height=7.25cm]{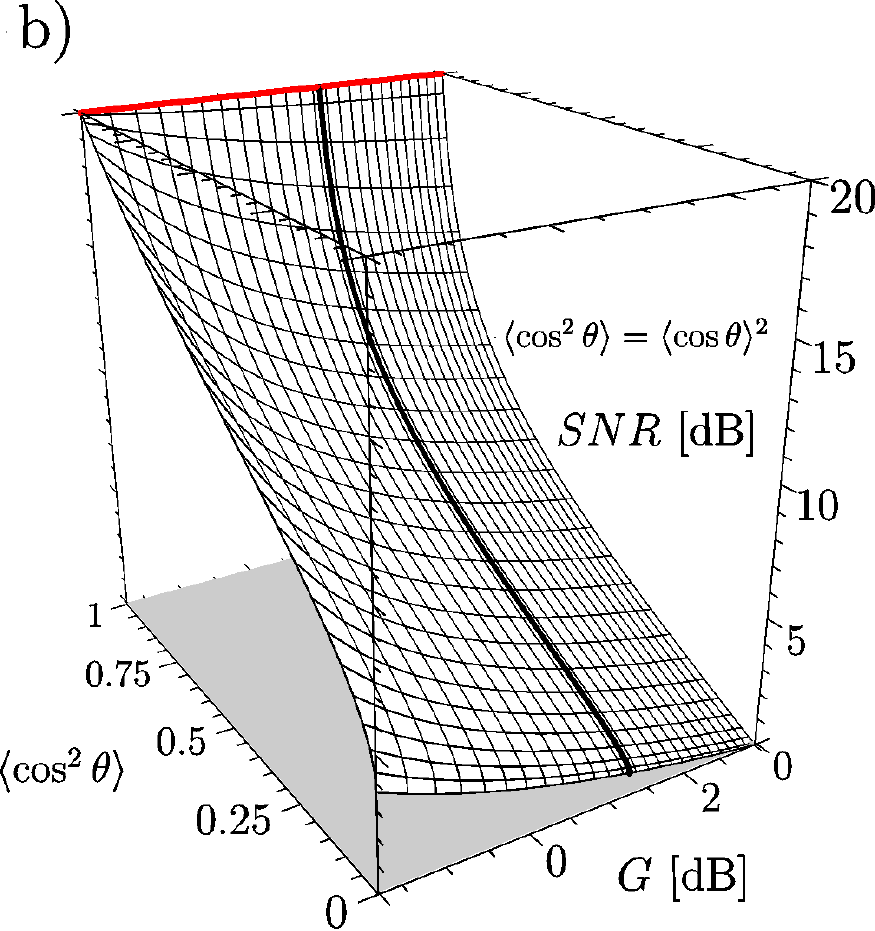}
\caption{\label{fig:4} (Color online) Signal to noise ratio $SRN$ versus gain
$G$ for a given input qubit distribution quantified by $\langle\cos^{n}\theta\rangle$
for $n=1,2$ fully characterizes the CBA. (a):
we assumed distributions with $\langle\cos\theta\rangle=0$; if $\langle\cos^{2}\theta\rangle=1/3$,
the amplifier is universal; if $\langle\cos^{2}\theta\rangle=0$ the
amplifier is phase-covariant optimized for equatorial qubits. If $\langle\cos^{2}\theta\rangle=1$
(red curve) then the signal can be amplified classically with infinite
$SNR$ ratio because the states of $\cos\theta=\pm1$ can be discriminated
deterministically. (b): if the state of the qubit is known
\textit{a priori}, e.g. $\cos\theta=\pm1$, $\langle\cos^{2}\theta\rangle=\langle\cos\theta\rangle^{2}$,
the qubit can be amplified with an arbitrarily high gain (red curve).
Solid black curves correspond to experimentally demonstrated amplification
discussed in the text below.}
\vskip-0.5cm 
\end{figure}

The efficiency $P$ can be understood as the success probability 
$P$ of  a copying process $\rho\to\rho'$ providing two copies $\rho'$ of the original qubit
$\rho=|\Psi\rangle\langle\Psi|$ for $\ket{\Psi}=\cos(\theta/2)\ket{H} + e^{i\phi}\sin(\theta/2)\ket{V}$ with fidelity $\mathcal{F}$, so
that $\rho'=(2\mathcal{F}-1)\rho+(1-\mathcal{F})\openone$ and $\openone/2$
is a maximally mixed state. The intrinsic noise of this
operation quantified by inverted signal to noise ratio ($SNR$) 
decreases along with increasing fidelity $\mathcal{F}$. The relation
between the $SNR$ and fidelity is:
\begin{equation}
SNR(\mathcal{F})=\tfrac{x}{1-x},
\end{equation}
where $x=2\mathcal{F}-1$. In perfect copying $\mathcal{F}=1$ and
the $SNR\to\infty$. For the best classical cloning of an unknown qubit 
$SNR=1$ (0\,dB) for $\mathcal{F}=0.75$. Thus, we will use this $SNR$
value as a threshold for quantum amplification.
Figure~\ref{fig:4} depicts the $SNR$ as a function
of gain and specific qubit distributions. Now, we can express the condition on increasing product state capacity for depolarizing channel defined in Eq.~(\ref{eq:psc}) as
\begin{equation}
G>1+SNR^{-1},
\end{equation}
where $G=2P$, i.e., for $\eta\ll1$.

\subsection{Applying hybrid quantum-classical cloning}

As we know from the previous section, for small values of $\eta\ll 1$ the gain of the CBA reads $G\equiv G^{(\mathrm{CBA})}_{T}=2P$, where $P$ can be arbitrarily close to $1$. However, in linear optical
quantum cloning experiments the value of $P$ is lower. The CBA performs 
optimal amplification of qubits
whose distribution on the Bloch sphere is axially symmetric. The quantum
cloning strategy in~\cite{Bartkiewicz10} is optimal for
these distributions, allowing for the copying fidelity to exceed $\mathcal{F}=5/6$ (universal cloning limit). However, the process is inherently probabilistic and cannot be directly
employed for amplification: it produces two good copies with the probability
{$P_{A}<1/2$}.

Fortunately, it is also possible to find a classical deterministic
strategy that uses \textit{a priori} knowledge regarding the state
distribution to be cloned. It randomly swaps the original qubit with the mean
state of the cloned distribution (the central state)~\cite{cloning}.
The CBA can combine those two optimal strategies to amplify photonic
qubits with a high probability $P$. The success rate of the CBA equals
\begin{equation}
P(\epsilon)=\epsilon+(1-\epsilon)P_{A}
\end{equation}
where $\epsilon$ describes mixing the quantum ($\epsilon=0$)
and the classical approach ($\epsilon=1$). For  axially-symmetric
cloning the success probability $P_{A}$ equals
\begin{eqnarray}
\nonumber
{P_{A}} & = & \tfrac{1}{6\cos^{2}\alpha_{+}}\big(\cos^{2}\tfrac{\theta}{2}\cos^{2}\alpha_{+}+\sin^{2}\tfrac{\theta}{2}\sin^{2}\alpha_{-}\big)\\
 &  & +\tfrac{1}{6\cos^{2}\alpha_{-}}\big(\sin^{2}\tfrac{\theta}{2}\cos^{2}\alpha_{-}+\cos^{2}\tfrac{\theta}{2}\sin^{2}\alpha_{+}\big).
\end{eqnarray}
The result is optimal if $\cos^{2}\alpha_{\mp}\geq\sin^{2}\alpha_{\pm}$.
The values of $\alpha_{\pm}$ depend on the shape of the qubit distribution 
$g(\rho)$~\cite{Bartkiewicz10}. $P(\epsilon)$ can be arbitrarily close to one 
(deterministic regime), providing maximal gain $G$  for $\epsilon\simeq1$ and
$\eta\ll1$, $G=2P(\epsilon)\simeq2$. Note that $P_{A}$
describes a general class of transformations including universal~\cite{Hillery},
phase-covariant~\cite{Bruss2000} and mirror phase-covariant 
cloners~\cite{Bartkiewicz09}.

Any spherical axisymmetric distribution  function
$g(\rho)=g(\theta)$ can be expressed by the Legendre polynomials
$P_n(\cos{\theta})$ as~\cite{Kaplan}:
\begin{eqnarray}
g(\theta) &=& \frac{1}{4\pi} \sum_{n=0}^{\infty} (2n+1)a_n P_n(\cos{\theta}),\\
a_n  &=& \int_{0}^{2\pi}\int_{-1}^{1}
g(\theta)P_n(\cos{\theta})\,{\rm d}\cos{\theta}\,{\rm d}\phi.
\end{eqnarray}
Therefore, any results depending on $g(\rho)$ can be expressed in terms of $a_n$.
For a normalized qubit distribution ($a_0=1$) one needs to know only $a_1$ and
$a_2$ in order to fully characterize the corresponding optimal
cloning transformation. For example, the values representing angles $\alpha_\pm$  
depend on a single parameter
\begin{equation}
\Gamma = \frac{6\sqrt{2}a_1(a_2-1)}{x_+x_-},
\end{equation}
where $x_\pm = 1+2a_2\pm3a_1$. As long as $|\Gamma| < 1$, we can express
the $\alpha_\pm$ angles as
\begin{eqnarray}
2\alpha_{\pm} =  \arcsin{\Omega} \pm \arcsin{\Gamma},
\end{eqnarray}
where
\begin{equation}
\Omega = \frac{2\sqrt{2}(1+2a_2)(1-a_2)}
{\sqrt{3x_+x_-(3+4a_2^2-3a_1^2-4a_2)}}. 
\end{equation}
However, for $|\Gamma|> 1$ there are two possibilities, i.e., 
$\alpha_+ = 0$ and $\alpha_ -= \frac{\pi}{2}$ or vice versa.
The corresponding optimal cloning process in this case is 
the phase covariant cloning of  Fiur\'a\v{s}ek ~\cite{Fiurasek2003}. 
The case of $a_1=0$ includes the equatorial phase covariant cloning for $\theta=\pi/2$ and the mirror phase covariant cloning~\cite{Bartkiewicz09}
($\alpha_+=\alpha_-$). If $a_1=a_2=0$, we obtain the universal cloning
transformation~\cite{Hillery}. The difference between the universal CBA using deteministic quantum cloning and our hybrid approach (constrained by linear optics) is described quantitively in Tab.~\ref{tab:1}.

\begin{table}
\caption{\label{tab:1}Comparison of the parameters of CBA using deterministic quantum cloning CBA${}_{\mathrm{DQC}}$ and   CBA${}_{\mathrm{LO}}$($\epsilon=0.5$) implemented within the framework of linear optics and using
hybrid quantum-classical cloning. The data suggests that even for a large amount of classical cloning
the CBA${}_{\mathrm{LO}}$($\epsilon=0.5$) provides amplification (gain $G>1$) and produces signal 
that is better in quality than the best classical amplifier ($F>0.75$). The total transmissivity of the channel is $\eta=\eta_A\eta_B=0.01$.}
\begin{ruledtabular}
\begin{tabular}{l c c c c c}
Amplifier 	& $\eta_A$ & $\eta_B$ & $G$ & $P$ & $\mathcal{F}$\\

CBA${}_{\mathrm{DQC}}$				& 0.01 & 1.00 & 1.00 & 1.00 & 0.83\\
			& 0.50 & 0.02 & 1.98 & 1.00 & 0.83\\
			& 1.00 & 0.01 & 1.99 & 1.00 & 0.83\\

CBA${}_{\mathrm{LO}}$($\epsilon=0.5$) 	 			 &  0.01 & 1.00 & 0.62 & 0.62 & 0.77\\
			 &  0.50 & 0.02 & 1.23 & 0.62 & 0.77\\
			 &  1.00 & 0.01 & 1.23 & 0.62 & 0.77\\
\end{tabular} 
\end{ruledtabular}
\end{table}


The average fidelity $\mathcal{F}$ of the clones $\rho'$ produced
by the CBA equals the weighted average over the Poincaré sphere
calculated using a measure $\mathrm{d}\omega={\rm d}\cos{\theta}\,{\rm d}\phi$
(the Haar measure) suitable for the spherical geometry of the
original qubits $\rho = \ket{\Psi}\bra{\Psi}$ multiplied by the distribution~$g(\rho)$,
i.e., $\mathcal{F}=\int\mathrm{d}\omega\, g(\rho)\mathrm{Tr}\{\rho\rho'\}$,
where $\rho'=(1-\epsilon)\rho_{q}+\epsilon\rho_{c}$, $\rho_{q}$
is the best quantum copy, and $\rho_{c}=(\sigma+\rho)/2$ is a state
obtained by the classical strategy for $\sigma$ being the best replacement
for $\rho$. This expression is maximized if $\sigma=\int\mathrm{d}\omega\,\rho g(\rho)$. 
For axially distributed input qubits $|\Psi\rangle$ we obtain 
\begin{equation}
\mathcal{F}=(1-\varepsilon)\mathcal{F}_{\mathrm{A}}+\tfrac{\varepsilon}{4}(3+\langle\cos\theta\rangle^{2})
\end{equation}
where $\varepsilon = \epsilon/P(\epsilon)$, $\sigma=(\openone+\langle\cos\theta\rangle\sigma_{z})/2$, and
\begin{eqnarray}
\mathcal{F}_{\mathrm{A}} & = & \tfrac{1}{8}[2(3+\cos2\alpha_{+})\langle\cos^{4}\tfrac{\theta}{2}\rangle+2(3+\cos2\alpha_{-})\langle\sin^{4}\tfrac{\theta}{2}\rangle\nonumber \\
 &  & +(\sin^{2}\alpha_{+}+\sin^{2}\alpha_{-}+2\sqrt{2}\sin\Omega)\langle\sin^{2}\theta\rangle],
\end{eqnarray}
is quantum cloning fidelity with $\Omega=\alpha_{+}+\alpha_{-}$~\cite{Bartkiewicz10}. For 
the universal CBA one obtains $\sin\alpha_{\pm}=1/\sqrt{3}$ and
$\mathcal{F}_{\mathrm{A}}=5/6$ resulting in $\mathcal{F}^{\mathrm{u}}=(1-\varepsilon)5/6+3\varepsilon/4$.

To streamline our experiment we focused on axially-symmetric input
qubit distributions with $\langle\cos\theta\rangle=0$  [see  Fig.~\ref{fig:4}(a)]
and $\langle\cos\theta\rangle^2=\langle\cos^2\theta\rangle$ [see  Fig.~\ref{fig:4}(b)]. They encompass mirror phase-covariant, phase-covariant and universal cloners. 
If the mirror symmetry is slightly broken, the optimal QCM becomes a phase-covariant
cloner~\cite{Bartkiewicz10} optimized for $\langle\cos^{2}\theta\rangle=\langle\cos\theta\rangle^{2}$,
i.e. $\cos\alpha_{\pm}=1/\sqrt{2}$ and $G(\epsilon)=2(1-\epsilon)/3+2\epsilon$.

\section{Comparison with heralded amplifier}

Both the HA and CBA are called ``amplifiers'' and are 
used to increase the probability of detecting a photon. 
Their principle of operation is very different. The HA
increases this probability by rejecting cases when a photon is not likely 
to appear. The CBA does that by sending more photons.
However, despite their differences the HA and the CBA have a lot in common.
The term ``amplifier'' implies that there is some gain (in energy)
and possibly some noise added. This is true for both cases,
if we consider the number of photons received in a time window.
For the CBA this window is just an arbitrary time interval. 
For the HA this time interval is the time over which the HA 
heralds the arriving photons (the remaining time can be used
for some other purpose). It is natural to ask whether one of these
amplifiers gives better results.
  
Here we will demonstrate that the two amplifiers perform complementary tasks.
Depending on the amplification regime, one of the amplifiers becomes
more useful than the other. We will assume that both the amplifiers
are optimized for a uniform qubit distribution over the Bloch sphere.
The results of our study are sumarized in Tab.~\ref{tab:2}.

\begin{table}
\caption{\label{tab:2} Comparison of the parameter of HA form Ref.~\cite{Evan13} (it provides higher success probability than the scheme proposed in Ref.~\cite{gisin10ampl}) with the CBA${}_{\mathrm{LO}}$($\epsilon=0.5$) implemented within the framework of linear optics and using hybrid quantum-classical cloning for $\epsilon = 0.5$. The paramteter $r=0.294$ of the HA represents reflectivity of beamsplitters that is set so that gain of the amplifier will reach the one of the CBA for $\eta_A = 0.5$. If gain of both the amplifiers is the same, then CBA has superior success rate. However, CBA provides lower signal fideliy than the noiseless amplifier. The total transmissivity of the channel is $\eta=\eta_A\eta_B=0.01$. The meaning of parameters $\eta_A$ and $\eta_B$ is explained in Fig.~\ref{fig:concept}.}
\begin{ruledtabular}
\begin{tabular}{lccccc}
Amplifier 			& $\eta_A$ & $\eta_B$ & $G$ & $P$ & $\mathcal{F}$\\

CBA${}_{\mathrm{LO}}$($\epsilon=0.5$) 	
			  &  0.01 & 1.00 & 0.62 & 0.62 & 0.77\\
			  &  0.50 & 0.02 & 1.23 & 0.62 & 0.77\\
			  &  1.00 & 0.01 & 1.23 & 0.62 & 0.77\\

HA($r=0.294$)  & 0.01 & 1.00 & 1.58 & 0.09 & 1.00 \\
	   		   & 0.50 & 0.02 & 1.23 & 0.11 & 1.00 \\
			   & 1.00 & 0.01 & 1.00 & 0.14 & 1.00\\
\end{tabular} 
\end{ruledtabular}
\end{table}

\subsection{Postamplification}

When both the CBA and the HA are used at the end of the channel, the HA implements
the following transformation~\cite{gisin10ampl,Ralph10}
\begin{equation}
\rho_{\mathrm{in}}'\to\rho_{\mathrm{out}}=(1-P)\rho_{\mathrm{vac}} + 
P\rho_{\mathrm{out}}',
\end{equation} 
where the input state is the attenuated qubit $\rho_{\mathrm{in}}' 
=\eta\rho_{\mathrm{in}} + (1-\eta_A)\rho_{\mathrm{vac}}$ and $P$ is the success 
probability of the device. 
With probability $P$ one can herald the state  $\rho_{\mathrm{out}}' = \tfrac{1}{N}
[(1-\eta_B)\rho_{\mathrm{vac}} + g^2\eta_B\rho]$, where $N = 1-\eta_A + \eta_A g^2$.
The nominal gain of the HA is defined as $G^{(\mathrm{HA})}_{\mathrm{nom}} = 
g^2/N$. Nevertheless, the transmission gain of the HA has to include the 
probability of the successful operation of the device and is defined as
\begin{equation}
G^{(\mathrm{HA})}_T = P G^{(\mathrm{HA})}_{\mathrm{nom}} \leq 1,
\end{equation} 
where the inequality can be  saturated only for $\eta_A=1$. However, if we are interested only in the cases when the arrival of the signal has been announced, the transmission gain reads
\begin{equation}
G^{(\mathrm{HA})}_{T'} = G^{(\mathrm{HA})}_{\mathrm{nom}} > 1,
\end{equation} 
The experimentally observed success probability values are low, e.g., $P=0.05$ for $G^{(\mathrm{HA})}_{\mathrm{nom}}=3.3\pm 0.6$ 
in \cite{Ralph10}. 
 
In comparison, if the input of the CBA is a damped photon in state
$\rho_{\mathrm{in}}'=\eta_A\rho_{\mathrm{in}} + (1-\eta_A)\rho_{\mathrm{vac}}$ and 
after the amplification process the photon is not damped, the transmission gain 
equals
\begin{equation}
G^{(\mathrm{CBA})}_T =  P \leq 1.
\end{equation} 
This means that the amplification gain defined as a ratio of probabilities of finding 
at least one photon at the output with and without amplification [see 
Eq.~(\ref{eq:gain})] does not reveal the apparent increase in signal intensity caused
by duplicating the signal. Using an alternative definition of 
transmission gain based on the ratio of the mean photon numbers with and 
without amplification would reveal  that the signal intensity grows depending on 
$P$.

The fair comparison of the two amplifiers should be performed for 
gains $G^{(\mathrm{CBA})}_T$ and $G^{(\mathrm{HA})}_{T'}$. Thus, in this scenario none of the amplifiers are useful for increasing the traditionally defined transmission rate (average transmission speed over the whole time of transmission). However, when the transmission gain for the HA is calculated only for the cases when the arrival of the photon is announced (average transmission speed over the time over which the signals were announced), the HA provides genuine amplification while CBA introduces losses and noise (for universal cloning qubit fidelity is $\mathcal{F}\leq 5/6$). Placing the HA at the end of the line (where we have a mixture of vacuum and the initial signal) can noiselessly increase the probability of finding the heralded photon to one (i.e. infinite nominal gain), at the expense of lowering the probability of announcing the arriving signal (this is by design impossible with the CBA). The CBA becomes useful if there is some more damping ahead. 

\subsection{Preamplification}
For a fixed success rate $P$, the transmission gain of the CBA is twice as high as 
that of the HA, which can be seen if the output states of the two devices are compared 
for a low channel 
transmittance $\eta_B\ll1$. The probability of finding at least one photon at
the output of the CBA amplifier is $P^{(\mathrm{CBA})}_{n>0}=2P\eta_B$, whereas 
for the HA it reads $P^{(\mathrm{HA})}_{n>0}=P\eta_B$. If none of the amplifiers is used 
$P^{(0)}_{n>0}=\eta_B$. This results in $G^{(\mathrm{CBA})}_{T}\approx G=2P>1$, if 
$P>1/2$, for the CBA and $G^{(\mathrm{HA})}_T=P\le1$ (or $G^{(\mathrm{HA})}_{T'}=1$) for the HA.

If we consider qubit subspace fidelity, then for the  CBA it is equal $\mathcal{F}$ and for the HA it equals $1$. Thus, the HA performs better with regard to this figure of merit. Another important quantity for amplifiers is the output fidelity (as defined in Ref.~\cite{Ralph10}) defined for $\rho_{\mathrm{in}}=\rho$ and a single-mode output of the CBA as
$F^{(\mathrm{output,1})}_{\mathrm{CBA}} = \mathrm{Tr}(\rho_{\mathrm{out,1}}\rho)=P\eta_B\mathcal{F}$,
where  $\rho_{\mathrm{out,1}}$ 
is the reduced single-copy density matrix of the first output (assuming that all the  copies are the same) including $\rho'$ and some vacuum.  To perform a fair comparison we should sum contributions from all the output modes. In the limit $\eta_B\ll1$ the modes are independent because there is only a small chance of receiving both copies. If we now physically combine all output modes of the CBA we will obtain (for $\eta\ll1$ and $1\to 2$ cloning) 
\begin{equation}
F^{(\mathrm{output})}_{\mathrm{CBA}} =N_{\mathrm{copy}} \mathrm{Tr}(\rho_{\mathrm{out,1}}\rho)= 2\eta_B P \mathcal{F}.
\end{equation}
For a single-mode output of the HA the output fidelity becomes
\begin{equation}
F^{(\mathrm{output})}_{\mathrm{HA}} = \mathrm{Tr}(\rho_{\mathrm{out}}\rho) = \eta_B P.
\end{equation}
This means that for a fixed success rate $P$ and number of copies 
$N_{\mathrm{copy}}=2$ the CBA is better than HA as long as 
$2\mathcal{F} > 1$, which is always true. If the fidelity $\mathcal{F}$ of the CBA 
is optimized so that $2P\mathcal{F}>1$, the CBA outperforms the HA when both 
amplify qubits at the  beginning of the transmission line. Leaving the qubit 
unaltered is preferable to probabilistic heralding of its presence, even if the 
success probability of the HA equals 1.

\section{The experiment}

In our experiment we measure only $\rho_{1,2}$ ($\rho'$) and $P$
(see Fig.~\ref{fig:3}), but these are enough to estimate $G$ 
(for $\eta \ll 1$) and $SNR$. For technical reasons we were not 
able to separate the unsuccessful quantum cloning events from 
other cases by  means other
than the  postselection on the successful detection of $\rho_{1,2}$.
Thus, our proof-of-principle experiment investigates 
channel-independent properties of a linear-optical CBA
(as in the case of $\eta \ll 1$).

\begin{figure}
\includegraphics[width=8cm]{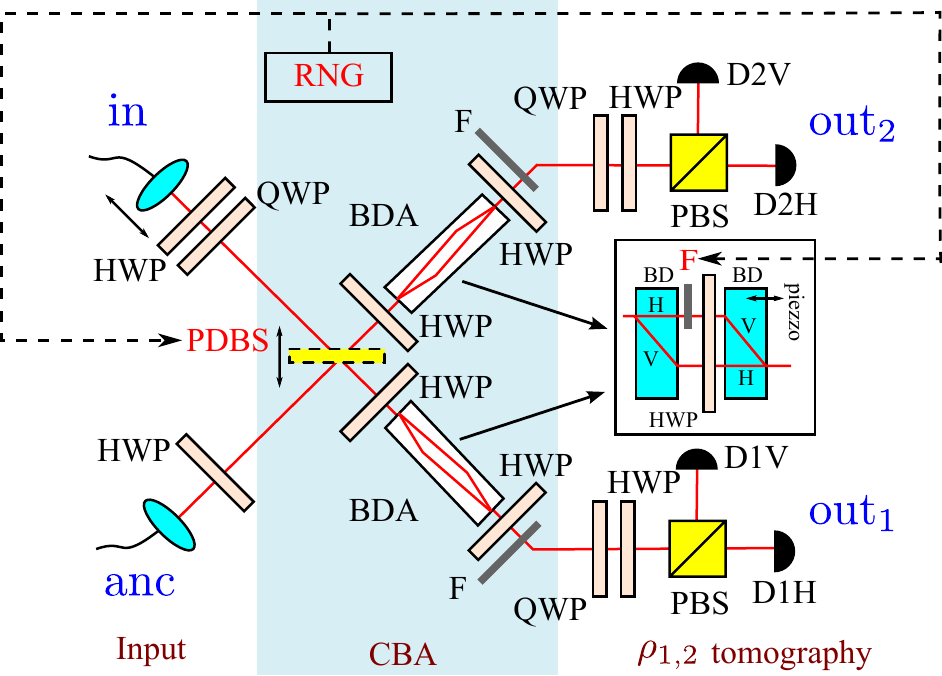} 
\caption{\label{fig:5} (Color online) Implementation of the cloning-based
quantum amplifier (CBA). A random number generator RNG provides a number 
$r\in[0,1]$ and switches the device between its two regimes
of work:  optimal quantum cloning (if $r \leq 1-\epsilon$) and the mixing of
input qubit $\rho$ with the central state $\sigma$ (if $r>1-\epsilon$). The latter 
regime requires removing the PDBS and filters F from BDAs. PDBS, polarization 
dependent beam splitter; BD, beam displacer; BS, beam splitter; HWP
(QWP), half (quarter) wave plate; F, neutral density filter. After the signal $\rho$ and 
ancillary photons are prepared (by means of HWPs and a QWP) they interact on a PDBS. 
Next, if the photons exit at separate ports, a specific polarization component (set by 
HWPs) of both photons is attenuated in the respective BDAs (another pair of HWPs 
restores the original polarization). Just before exiting the CBA the intensity in both 
output ports is balanced by a pair of neutral density filters F. Finally, a polarization 
analysis is performed.}
\end{figure}

\subsection{The source}

Type I degenerate spontaneous parametric down-conversion
(1\,cm long LiIO$_{3}$ crystal) generates pairs of photons used
as input states. The crystal is pumped by a 413\,nm continuous
wave Kr$^{+}$ laser (200\,mW). Half- (HWP) and quarter wave
plates (QWP) encode the qubits in a photonic polarization. 
The signal photon (\textquotedbl{}in\textquotedbl{}) is further
amplified, the idler (\textquotedbl{}anc\textquotedbl{})
becomes an ancillary photon whose state depends on the
type of cloning used in the CBA. Figure~\ref{fig:5} shows the
implementation setup of the CBA.

\subsection{Optimal quantum cloning}

The first strategy is quantum-based ($\epsilon=0$). The device functions
as a phase-covariant and mirror phase-covariant cloner~\cite{Lemr12}.
The optimal cloning transformation can be now written in the form of a unitary
transformation
\begin{subequations}
\begin{eqnarray}
\ket{HHH} &\rightarrow & \cos\alpha_+\ket{HHV} +
\sin\alpha_+\ket{\psi_+}\ket{H},\\
\ket{VHH} & \rightarrow & \cos{\alpha_-}\ket{VVH} +
\sin{\alpha_-}\ket{\psi_+}\ket{V},
\end{eqnarray}
\end{subequations}
where $\ket{\psi_+}=\left(\ket{HV} + \ket{VH} \right)/\sqrt{2}$.
The first mode is the signal mode, the last two correspond to
ancillary modes. After the transformation the clones are encoded
in the last two modes and the third mode is removed. 
The unitary transformation could be implemented in a deterministic
way in a properly tailored system.
However, being limited by linear optics, in our experiment
we perform an equivalent stochastic operation
consisting of two operations involving only two-photon interactions
\begin{subequations}
\begin{eqnarray}
\ket{HH}_{\mathrm{in,anc}} &\rightarrow & \cos\alpha_+\ket{HH}_{1,2}  ,\\
\ket{VH}_{\mathrm{in,anc}}  & \rightarrow & \sin{\alpha_-}\ket{\psi_+}_{1,2} ,
\end{eqnarray}
\end{subequations}
and
\begin{subequations}
\begin{eqnarray}
\ket{HV}_{\mathrm{in,anc}}  &\rightarrow & \sin\alpha_+\ket{\psi_+}_{1,2},\\
\ket{VV}_{\mathrm{in,anc}}  & \rightarrow & \cos{\alpha_-}\ket{VV}_{1,2}.
\end{eqnarray}
\end{subequations}
The transformations are probabilistic and they work with the success rate $P_A$.
The ancillary photon is polarized either horizontally or vertically.
The signal and ancillary photons interfere on a polarization dependent
beam splitter (PDBS). 

\begin{figure}
\includegraphics[width=8cm]{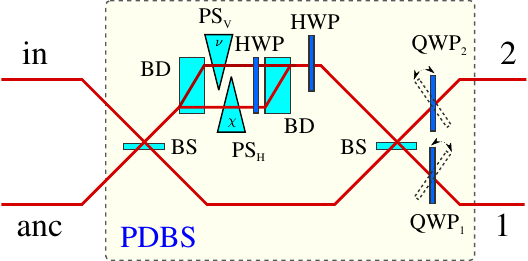} 
\caption{\label{fig:6} (Color online) Implementation of polarization
dependent beam splitter PDBS: BD, beam displacer; BS, symmetric beam splitter; HWP
(QWP${}_{1,2}$), half (quarter) wave plate; PS${}_{H(V)}$, phase shifter. By manipulating the phase shifts $\nu$ and $\chi$ of H and V polarization components respectively, the interferometer acts as a PDBS with variable reflection and transmittance for each of the H and V polarizations. To remove the phase difference $\nu-\chi$ between the phases of V and H polarization component of both output modes, we use QWP${}_1$ and QWP${}_2$ (or Pockels cells) set to shift phases of V-polarized photons by $\chi-\nu$. It can be verified that the setup output may be expressed in terms of annihilation operators as $a_{2,s} =  \sqrt{\eta_s} a_{\mathrm{anc},s} - \sqrt{1-\eta_s} a_{\mathrm{in},s}$, $a_{1,s} = \sqrt{\eta_s} a_{\mathrm{in},s} +\sqrt{1-\eta_s}a_{\mathrm{anc},s}  $, where $\eta_s=\cos^2\chi,\cos^2\nu$ for $s = H,V$, respectively. If there is no phase shift, the PDBS acts as if it was removed. 
}
\end{figure}

Repeated scanning of the Hong-Ou-Mandel dip ensures interferometric stability. 
This is achieved by shifting the PDBS mounted on a motorized translation stage and
fixing it in the optimal position. The PDBS  implemented as  a Mach Zehnder 
interferometer  including conventional beam splitters (BS), phase shifters (PS), a wave 
plate and beam dividers (BD) is shown in  Fig.~\ref{fig:6}.
Overall PDBS transmittivities obtained experimentally equal
$\eta_H=76\%$ and $\eta_V=18\%$ for horizontal and vertical polarizations, respectively.
To maximize the output state fidelity polarization dependent
filtration is applied in both output ports of the PDBS by two blocks 
(see BDA blocks in Fig.~\ref{fig:5}) consisting of
two BDs splitting and subsequently merging horizontal
and vertical polarizations, and a neutral density filter placed between
them introducing polarization dependent losses. Filtering
corrects beam splitter imperfections at the expense
of success probability $P_A$. It also allows us to optimize
the setup for various $\alpha_\pm$. If no filtering is applied,
the setup (consisting only of the PDBS) approximates a transformation 
corresponding to phase-covariant cloning with $P_A\approx 1/3$, i.e.,
\begin{subequations}
\begin{eqnarray}
\ket{HH}_{\mathrm{in,anc}} &\rightarrow & \ket{HH}_{1,2}  ,\\
\ket{VH}_{\mathrm{in,anc}}  & \rightarrow & \ket{\psi_+}_{1,2} ,
\end{eqnarray}
\end{subequations}
and
\begin{subequations}
\begin{eqnarray}
\ket{HV}_{\mathrm{in,anc}}  &\rightarrow & \ket{\psi_+}_{1,2},\\
\ket{VV}_{\mathrm{in,anc}}  & \rightarrow & \ket{VV}_{1,2}.
\end{eqnarray}
\end{subequations}
With probability $1-P_A$ the two photons bunch in one output
port, hence the cloning operation fails.
It is apparent that to introduce the scaling factors used for an arbitrary 
axially symmetric cloning we have to attenuate the H (V) polarization 
in both output modes by the same amount, depending on the state
of the ancillary photon. For more technical details on implementing
multifunctional quantum cloning see Refs.~\cite{Lemr12,Bartkiewicz12}. 

\subsection{Optimal classical cloning}

The second strategy constitutes classical copying.
It requires removing the PDBS and setting the ancillary state as the
central state $\sigma$ of the signal state distribution. The signal and ancillary
photons propagate without interference (all filters are removed). 
The signal and ancillary modes are swapped with probability $1/2$.  
Our setup permits to deterministically
prepare an arbitrary ancillary and signal state with almost
perfect fidelity with respect to the target state.

\subsection{Hybrid cloning}

A random number generator
(RNG) switches the device between two regimes of work, permitting
two different strategies (quantified by $\epsilon$)
described below. The RNG provides a random number $r\in [0,1]$ given by
a uniform distribution. If this number 
satisfies $r\leq 1-\epsilon$, the quantum strategy is implemented, 
else ($r>1-\epsilon$) the classical strategy is applied.

\subsection{Data analysis}

A complete two-photon polarization tomography is performed in both 
strategies \cite{halenkova2012detector}. To do that
we project the states at the two output ports out${}_{1(2)}$ to horizontal,
vertical, diagonal and anti-diagonal linear polarizations and right-
and left-hand circular polarizations. Time-bin encoding
can be used to deterministically combine the two output modes into
a single mode suitable for transmission through a single optical fiber.
Here the modes are separated for practical reasons. 
Photons are detected in both modes by the single photon counting
modules (detectors), and the numbers of coincidences per 5\,s intervals
are registered for all relevant two-mode polarizations. 
Typically, we accumulate about 100\,000 coincidences
for each output state. The maximum likelihood
method is used to reconstruct the photonic polarization density matrices \cite{Jezek03}.
The matrices are used to derive the experimental values of $SNR$ for a given gain $G=2P$.

To determine $P$ we compare the calibration coincidence rate
with that observed during the measurements. 
The former is measured in the setup with the PDBS
shifted out of its position so that the reflection is no longer coupled
to the detectors. All filters are retracted. This is how we distinguish 
between technological losses (caused by the
50\% single photon detection efficiency, imperfect
fiber coupling, and back-reflection) and the
fundamental success probability of the procedure~\cite{Lemr12}.

\begin{figure}
\includegraphics[width=7cm]{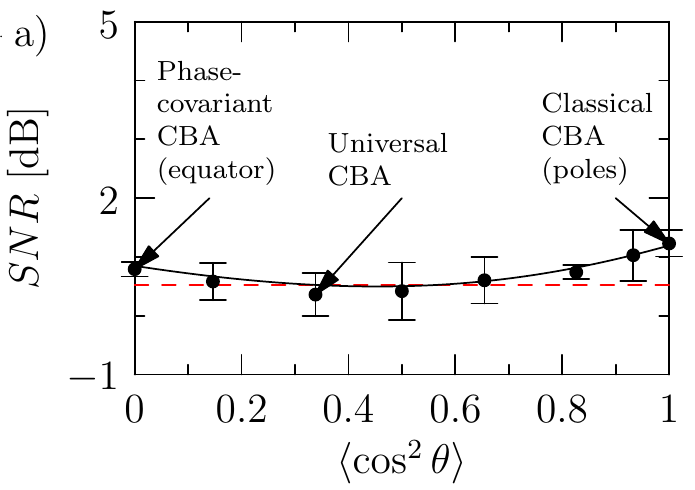} 
\includegraphics[width=7cm]{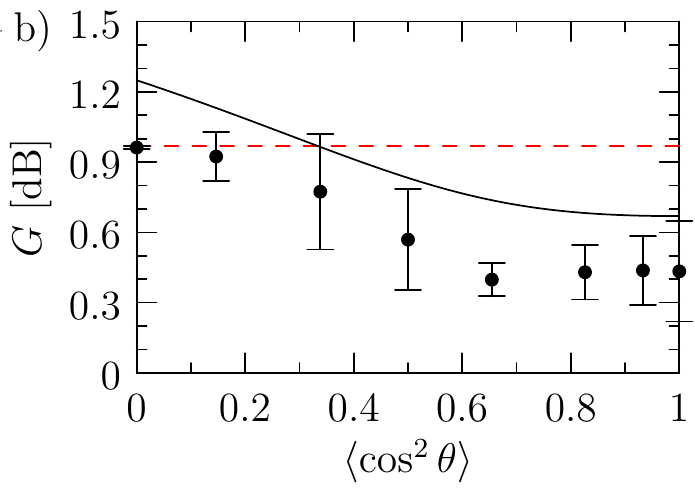} 
\caption{\label{fig:7} (Color online) Signal-to-noise ratio $SNR$ and amplification
gain $G$ measured for the cloning-based amplifier (CBA) with the
mixing parameter $\epsilon=1/2$ as a function of the input qubit
distribution on the Poincaré sphere, parametrized by $\langle\cos^{2}\theta\rangle$
with $\langle\cos\theta\rangle=0$, i.e., mirror phase-covariant
regime (see Fig.~\ref{fig:9} for $\langle\cos^{2}\theta\rangle=\langle\cos\theta\rangle^{2}$
regime). Dashed lines indicate the value of fidelity and gain for
the universal CBA. The details are given in the main text.}
\end{figure}

Figure~\ref{fig:7} depicts the $SNR$ ratio and gain $G$
measured for the CBA with the mixing parameter $\epsilon=1/2$ as
a function of the input qubit distribution on the Poincaré sphere,
parametrized by $\langle\cos^{2}\theta\rangle$, with $\langle\cos\theta\rangle=0$
(mirror phase-covariant distributions). The fidelity of the universal
CBA $\mathcal{F}_{\mathrm{exp}}^{\mathrm{u}}=0.758\pm0.018$ is, because of losses $\varepsilon_{\mathrm{th}}<\varepsilon_{\mathrm{ex}}$, smaller than its theoretical value $\mathcal{F}_{\mathrm{th}}^{\mathrm{u}}\approx0.77$.
These values constitute the lowest fidelities for the CBA in this
regime. However, both of these values are above the quantum amplification threshold for universal copying. 
The fidelities of the phase-covariant CBA for equatorial qubits
$\mathcal{F}_{\mathrm{exp}}^{\mathrm{pc}}=0.768\pm0.004$ and of the
classical (polar) CBA $\mathcal{F}_{\mathrm{exp}}^{\mathrm{c}}=0.773\pm0.012$
follow the trend of their theoretical values: $\mathcal{F}^{pc}_{\mathrm{th}}=0.78$,
$\mathcal{F}^{c}_{\mathrm{th}}=0.79$. The success rates for these three CBAs for
$\epsilon=1/2$ equal: $P_{\mathrm{th}}^{\mathrm{u}}=5/8$ ($P_{\mathrm{exp}}^{\mathrm{u}}=0.598\pm0.018$),
$P_{\mathrm{th}}^{\mathrm{pc}}= 2/3$ ($P_{\mathrm{exp}}^{\mathrm{pc}}= 0.624\pm0.005$),
$P_{\mathrm{th}}^{\mathrm{c}}= 7/12$ ($P_{\mathrm{exp}}^{\mathrm{c}}=0.553\pm0.014$).
Experimental results place below the theoretical curves for the
ideal setup components (solid) due to the imperfections of the PDBS.
The setup was optimized for $SNR$ which lowered the value
of gain $G$ by approx. 0.2\,dB. Note that the PDBS-related
imperfections are not included in the technological losses defined
above and, consequently, have an influence on the calculated success
probabilities. We also examined the performance of the device in
Fig.~\ref{fig:5} operating in the classical ($\epsilon=1$), quantum
($\epsilon=$0), and intermediate ($0<\epsilon<1$) regimes. 
The $SNR$ was measured as a function of $G$, parametrized by $\epsilon$.
As shown in in Fig.~\ref{fig:4}, two qubit distributions with $\langle\cos^{2}\theta\rangle=1/3$ and
$\langle\cos^{2}\theta\rangle=0$, both for $\langle\cos\theta\rangle=0$,
were given as input. The shaded areas of $G>0$\,dB and $SNR>0$\,dB show
where the device functions as a universal and phase-covariant CBA.
 
\begin{figure}
\includegraphics[width=7cm]{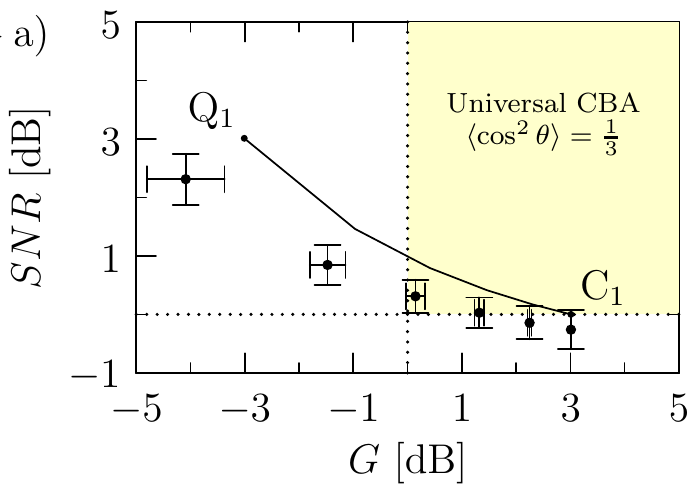} 
\includegraphics[width=7cm]{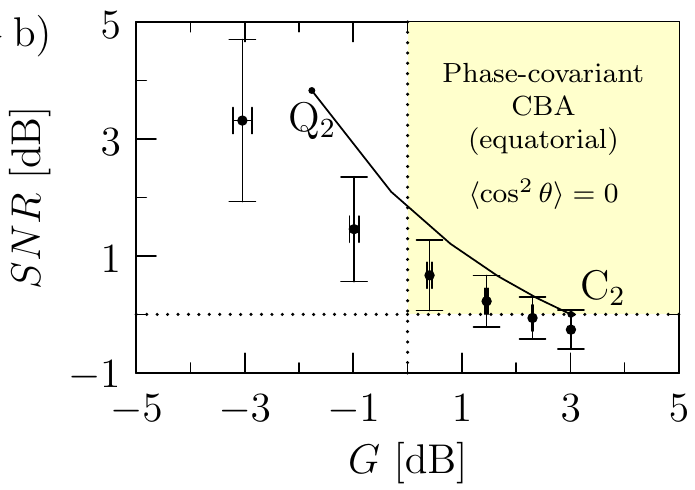} 
\caption{\label{fig:8} (Color online) $SNR$ as a function of gain $G$ measured
by the device from Fig.~\ref{fig:5} for input qubit distribution
$\langle\cos^{2}\theta\rangle=1/3$ (left) and $\langle\cos^{2}\theta\rangle=0$
(right) for $\langle\cos\theta\rangle=0$. Solid curves show theoretical
predictions for the ideal setup components, assuming the device operates
on a scale from purely quantum Q$_{1}=(-3.01,3.01)$ and Q$_{2}=(-1.76,3.83)$
($\epsilon=0$) to classical C$_{1(2)}=(3.01,0)$ ($\epsilon=1$)
regime. The marked areas of $G>0$\,dB and $SNR>0$\,dB highlight
the part of the regime where the device works as the universal and
phase-covariant CBA.}
\end{figure}

\begin{figure}
\includegraphics[width=7cm]{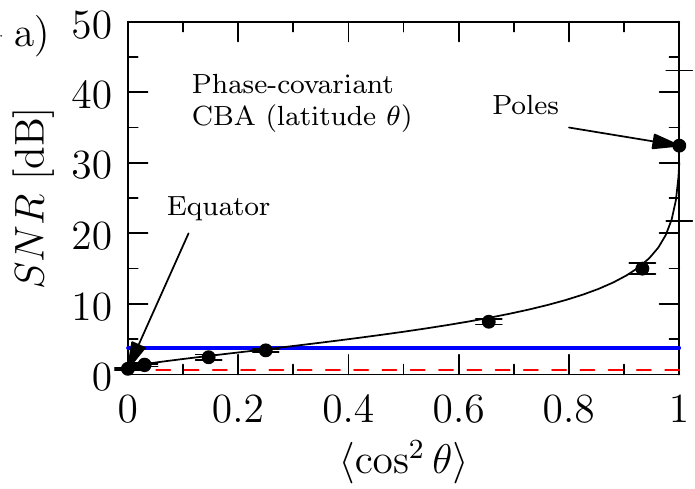}
\includegraphics[width=7cm]{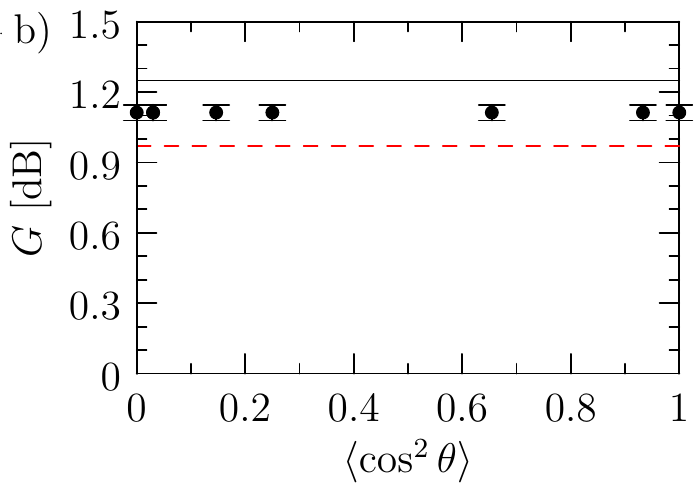}
\caption{\label{fig:9} (Color online) Signal-to-noise ratio $SNR$ and amplification gain $G$ measured for the cloning-based amplifier (CBA) with mixing parameter $\epsilon = 1/2$ as a function of the input qubit distribution on the Poincar\'e sphere, parametrized by $\langle\cos^2\theta\rangle$ with $\langle\cos^2\theta\rangle=\langle\cos\theta\rangle^2$, i.e., phase-covariant regime.
Red  dashed lines indicate the value of fidelity and gain for the universal CBA. Green dotted line indicates the $SNR$ above which the CBA provides improvement in product state capacity [see Eq.~(\ref{eq:psc})] for $\eta\ll 1$.}
\vspace{5mm}
\end{figure}

As shown in Fig.~\ref{fig:1},  there are two distinct regimes in which the cloning machine can operate.  The first one, described in detail in the paper, is the mirror phase-covariant regime, the other is the phase-covariat regime. The difference between these two regimes is that in the later case we have more \text{a priori} knowledge about the hemisphere from which the amplified qubits are selected. In the extreme case of the poles, cloning can be done in a perfect way since we know everything about the copied state. This is confirmed experimentally and the results are show in Fig.~\ref{fig:9}. This figure shows that with the increase of the \textit{a priori} knowledge about the states being copied we can improve the parameters of the amplifier in comparison with the universal case where the qubits to be amplified are evenly distributed over the Bloch sphere.

\section{Conclusions}

This paper presents a high-fidelity near-deterministic
genuine qubit amplifier (see Figs.~\ref{fig:7} and~\ref{fig:8}).
In contrast to heralded qubit amplifiers, our device produces pairs
of photons with $P>\frac{1}{2}$. In
this sense it increases the photon rate and achieves genuine amplification.
Because this effect is obtained at the expense of added noise we analyzed
the signal to noise ratio vs. the gain trade-off. The experimental
data fit the theoretical prediction, confirming amplifier
functionality. 
The CBA does not work properly if the cloning procedure fails by
sending both photons to out${}_1$ or out${}_2$ (see Fig.~\ref{fig:5}).
In our experiment we exclude such events by coincidence detection
but a full-field implementation will require other solutions 
(see, e.g., \cite{Everitt13} and references therein).

We compare the CBA working in a quantum regime ($\epsilon=0$)
to the stimulated emission process used for cloning. Since the available
nonlinearities in media are small, the efficiency of phase-covariant
cloning is very low: approx. 0.01 \cite{Simon00}. The efficiency
of the CBA equals 0.29, although the theoretical limit equals $\frac{1}{3}$. However, this limitation is platform dependent
and in other, more advanced implementations it could reach $1$.

Stimulated emission based amplification strongly depends on
the power of the amplified signal and is inefficient
if the average number of photons in the signal is too small.  In this regime
the CBA is potentially very useful, for the analysis of quantum channels with 
phase-independent damping as it extends the range of quantum communication.
In particular, we have demonstarated (see Fig.~\ref{fig:9}) that our implementaion
of CBA can yield values of $G$ and $SNR$ that are large enough
to increase product state capacity of the transmission channel described by Eq.~(\ref{eq:psc}).

\begin{acknowledgments}
 K.~B. gratefully acknowledges the support from
the Operational Program Research and Development for Innovations
-- European Regional Development Fund (project CZ.1.05/2.1.00/03.0058
and the Operational Program Education for Competitiveness - European
Social Fund (project CZ.1.07/2.3.00/20.0017 of the Ministry of Education,
Youth and Sports of the Czech Republic. K.~L. acknowledges support
by the Czech Science Foundation (Grant no. 13-31000P). This work was
also supported by the Polish National Science Centre under grant DEC-2011/03/B/ST2/01903.
M.~S. was supported by the EU 7FP Marie Curie Career Integration
Grant No. 322150 ``QCAT'', NCN grant No. 2012/04/M/ST2/00789, FNP
Homing Plus project No. HOMING PLUS/2012-5/12 and MNiSW co-financed
international project No. 2586/7.PR/2012/2. 
\end{acknowledgments}

\end{document}